\documentclass[final,3p,times]{elsarticle}

\usepackage{graphics}
\usepackage{graphicx}
\usepackage{epsfig}
\usepackage{amssymb}
\usepackage{color}

\journal{Turkish Journal of Physics}

\begin{document}
\begin{frontmatter}


\title{Dynamic phase transition features of the cylindrical nanowire driven by a propagating magnetic field}

\author[]{Erol Vatansever\corref{cor1}}
\cortext[cor1]{Corresponding author. Tel.: +90 3019547; fax: +90 2324534188.} \ead{erol.vatansever@deu.edu.tr}
\address{Department of Physics, Dokuz Eyl\"{u}l University, Tr-35160 \.{I}zmir, Turkey}



\begin{abstract}
Magnetic response of the spin-$1/2$ cylindrical nanowire to the
propagating magnetic field wave has been investigated by means of
Monte Carlo simulation method based on  Metropolis algorithm. The obtained
microscopic spin configurations
suggest that the studied system exhibits two types of dynamical
phases depending on the considered values of  system
parameters: Coherent  propagation of spin bands and spin-frozen or pinned phases,
as in the case of the conventional bulk systems under the influence of a
propagating magnetic field. By benefiting from the temperature dependencies of variances of
dynamic order parameter, internal energy and the derivative of dynamic order parameter
of the system, dynamic phase diagrams are also obtained
in related planes for varying values of the  wavelength of the propagating
magnetic field.  Our simulation results demonstrate that   as the  strength of the
field amplitude is increased,   the phase transition points   tend to shift
to the  relatively lower temperature regions. Moreover,
it has been observed  that dynamic  phase boundary line shrinks   inward when  the value of
wavelength of the external  field decreases.
\end{abstract}

\end{frontmatter}
\section{Introduction}\label{Introduction}
Interacting spin systems driven by a sinusoidally oscillating magnetic field can
exhibit distinctive  and fascinating dynamic behaviors, which can not occur for the corresponding
equilibrium spin systems. For the first time in Ref. \cite{Tome},  the authors applied their
theoretical model to investigate the physics underlying a simple ferromagnet being exposed to
a time dependent magnetic field. From their analysis,
the authors  concluded that amplitude and period of the oscillatory magnetic field play an important
role on the dynamic characters of the considered magnetic system. Since then, many theoretical \cite{Lo, Sides,
Buendia1, Buendia2,  Chakrabarti, Keskin, Shi, Park, Yuksel, Vatansever, Vatansever2,
Acharyya1, Acharyya2, Gallardo} and several
experimental works \cite{He, Robb, Suen, Berger, Riego} have been  carried to
understand the origin  of the  dynamic phase transitions. Based on the some of the previously published
studies, it is possible to say that there is a good consensus between dynamic phase
transitions and equilibrium phase transitions. For example, it has been reported that
the critical exponents of the two dimensional kinetic Ising model subjected to a
square-wave oscillatory magnetic field are consistent with the universality class of the
corresponding equilibrium Ising model \cite{Buendia2}. Moreover, it is recommended in these
references \cite{Vatansever2, Gallardo, Robb, Berger} that bias field appears to be a
conjugate field of the dynamic order parameter, which is time averaged magnetization
over a full cycle of the external field. In addition to the consensus in dynamic
phase transitions and equilibrium phase transitions, however, there are
inconsistencies in the literature, in view of the universality class of the
spin systems. For instance, recent detailed Monte Carlo (MC)
simulations studies show that there is a clear difference between critical dynamics of
the magnetic system with surfaces and its equilibrium case \cite{Park}. From their
finite-size scaling analysis,  the authors found that  nonequilibrium surface exponents
do not coincide with those of the equilibrium  critical surface. Very recently,
it is reported both experimentally  and theoretically that there are
metamagnetic fluctuations in the neighbourhood of  dynamic
phase transitions, which do not emerge in the thermodynamic behavior of
typical ferromagnets \cite{Riego}. Keeping these facts mentioned above in mind,
it is possible to say that more work is required to understand the origin of dynamic
phase transitions of magnetic systems driven by oscillatory magnetic field.

In addition to the works regarding the influences of an oscillating magnetic field (uniform
over space) on the dynamic characteristics  of the magnetic systems,
some efforts were  taken to investigate the dynamic features resulting
from standing and/or propagating magnetic field  waves \cite{Acharyya11, Acharyya12,
Acharyya13, Acharyya14}. In Ref. \cite{Acharyya11},  the authors applied their
theoretical model to elucidate the dynamical modes and nonequilibrium phase
transition of the spin-$1$ Blume-Capel model for two dimensional square lattice,
within the framework of MC simulation. It has been found
that, the wavelength of the external field plays an important role in the dynamic
nature of the considered system, in addition to the amplitude and period of the field.
The system presents two dynamical phases: Propagating spin wave and spin
frozen or pinned phases. In another interesting work, the physics behind a
simple ferromagnet  under the influence of a propagating magnetic field has
been discussed in detail \cite{Acharyya12}. It is shown that dynamic phase
boundary constructed in temperature versus applied field amplitude
tends to shrink as the strength of the wavelength of the external
field decreases. It is clear that particular attention in the works
mentioned above has only  been devoted to clarify the physics in the
bulk materials. To the best of our knowledge, there is no any
attempt to address the same problem for nanoparticles with surfaces, especially for
magnetic nanowires, driven by a propagating  or(and) standing magnetic field wave(s).
Magnetic nanowires are important materials and also potential  candidates  for applications in
advanced  nanotechnology including  magnetic  memory devices \cite{Irshad, Ivanov} and
biomedical  applications \cite{Ivanov2} due to
their own distinctive magnetic properties. For the sake of completeness, we would like to
emphasize that nonequilibrium dynamics as well as phase transition characteristics of
various types of magnetic nanowire systems under the existence of an oscillating field (uniform
over space) have been studied by employing
effective-field theory \cite{Deviren1, Deviren2, Kantar, Ertas} and also MC
simulation method \cite{Yuksel2, Yuksel3}.  From this point of view, we intend
to investigate the dynamic behaviors of  cylindrical nanowire system being subjected
to a  propagating magnetic field,  by utilizing MC simulation method. Our obtained results
indicate that the present system exhibits unusual and interesting magnetic dynamics
originating from the occurrence of  the propagating magnetic field, which have not been observed and
discussed so far, in view of the magnetic nanoparticles.

The outline of the remainder parts of the paper is as follows: In section \ref{Formulation},
we present the model and simulation details. The results and discussion are given in
section \ref{Results}, and finally section \ref{Conclusion} includes our conclusions.

\section{Model and Simulation Details}\label{Formulation}
The Hamiltoanian of the spin-$1/2$ cylindrical nanowire driven by a propagating magnetic field
can be written as follows:
\begin{equation}\label{Eq1}
\hat{H}=-J\sum_{\langle ij\rangle } S_{i}^z(x,y,z,t)S_{j}^z(x',y',z',t)-\sum_{i}h(x,y,z,t)S_{i}(x,y,z,t)
\end{equation}
here $S_{i}^z(x,y,z,t)=\pm 1$ is the Ising spin variable at any position $x, y$ and $z$ in the
nanowire. $J (>0)$ is the ferromagnetic spin-spin coupling term. The first summation
in Eq. (\ref{Eq1}) is over the nearest-neighbour spin couplings while the second one is over all
lattice sites in the system. $h(x,y,z,t)$ represents the propagating magnetic field,
which has the following:
\begin{equation}\label{Eq2}
h(x,y,z,t)=h_{0}\cos\left(\omega t-kz\right).
\end{equation}
For the sake of simplicity, the magnetic field is propagating
in the $z$ direction (long axis of the nanowire). Here, $h_{0}, \omega$ and $k$ refer to the amplitude,
angular frequency and  wavevector of the external field, and $t$ is time.
Period and wavelength of the magnetic field  are $\tau=2\pi/\omega$ and $\lambda=2\pi/k$, respectively.
It may be noted here that we have fixed the period of the external field as $\tau=100$
throughout the study.

By utilizing Metropolis algorithm \cite{Binder, Newman}, we employ MC simulation to understand
the magnetic nature of the cylindrical nanowire system driven by a propagating
magnetic field wave. The nanowire system is located on a simple cubic lattice, which has a length of $L_{z}=200$
and a radius $r=10$. For these selected values of the $L_{z}$ and $r$, total number
of the spins in the system is $N_{t}=63400$.  In order to mimic the cylindrical nanowire,
we follow the boundary conditions such that it is free  in $xy$ plane and periodic
along the $z$ direction of the system. The simulation procedure we follow in this study can be briefly
summarized as follows. The simulation starts from a high temperature $k_{B}T/J=5.0$ (here, $k_{B}$ and $T$
are Boltzmann constant and absolute temperature, respectively) using random
initial configurations, which corresponds to paramagnetic phase. Next,
it is slowly cooled  down until the temperature reaches  to the value
of $k_{B}T/J=10^{-2}$ with a relatively small  temperature step $k_{B}\Delta T/J=25\times 10^{-2}$.
In each temperature, $5\times10^4$ Monte Carlo steps per site (MCSS)  are discarded for thermalization
process, and next $5\times10^4$ MCSS are collected to determine  the thermal
variations of  physical  quantities used in this study. Numerical data were collected  over 20
independent samples.

\noindent We have calculated the instantaneous value of the magnetization at time $t$ as follows:
\begin{equation}\label{Eq3}
M(t)=\frac{1}{N_{t}}\sum_{i=1}^{N_t}S_{i}(x,y,z,t).
\end{equation}
Using $M(t)$, dynamic order parameter of the cylindrical nanowire system
can be defined as follows:
\begin{equation}\label{Eq4}
Q=\frac{1}{\tau}\oint M(t)dt.
\end{equation}
In order to detect the dynamic phase transition points of the present system, in addition to the derivative of
dynamic order parameter $dQ/dT$, we check the variance
of the $Q$ and internal energy $(E)$, which are defined as follows:
\begin{equation}\label{Eq5}
\chi_{Q}=N_{t}\left(\langle Q^2 \rangle-\langle Q \rangle^2 \right),
\end{equation}
and
\begin{equation}\label{Eq6}
\chi_{E}=N_{t}\left(\langle E^2 \rangle-\langle E \rangle^2 \right),
\end{equation}
here $E$ is the internal energy given in Eq. (\ref{Eq1}) over a full cycle of the
propagating magnetic field, which is defined in the following form:
\begin{equation}
E=-\frac{1}{\tau N_{t}}\oint \hat{H}dt.
\end{equation}

\section{Results and Discussions}\label{Results}
\begin{figure*}[!h]
\center
\includegraphics[height=4.cm]{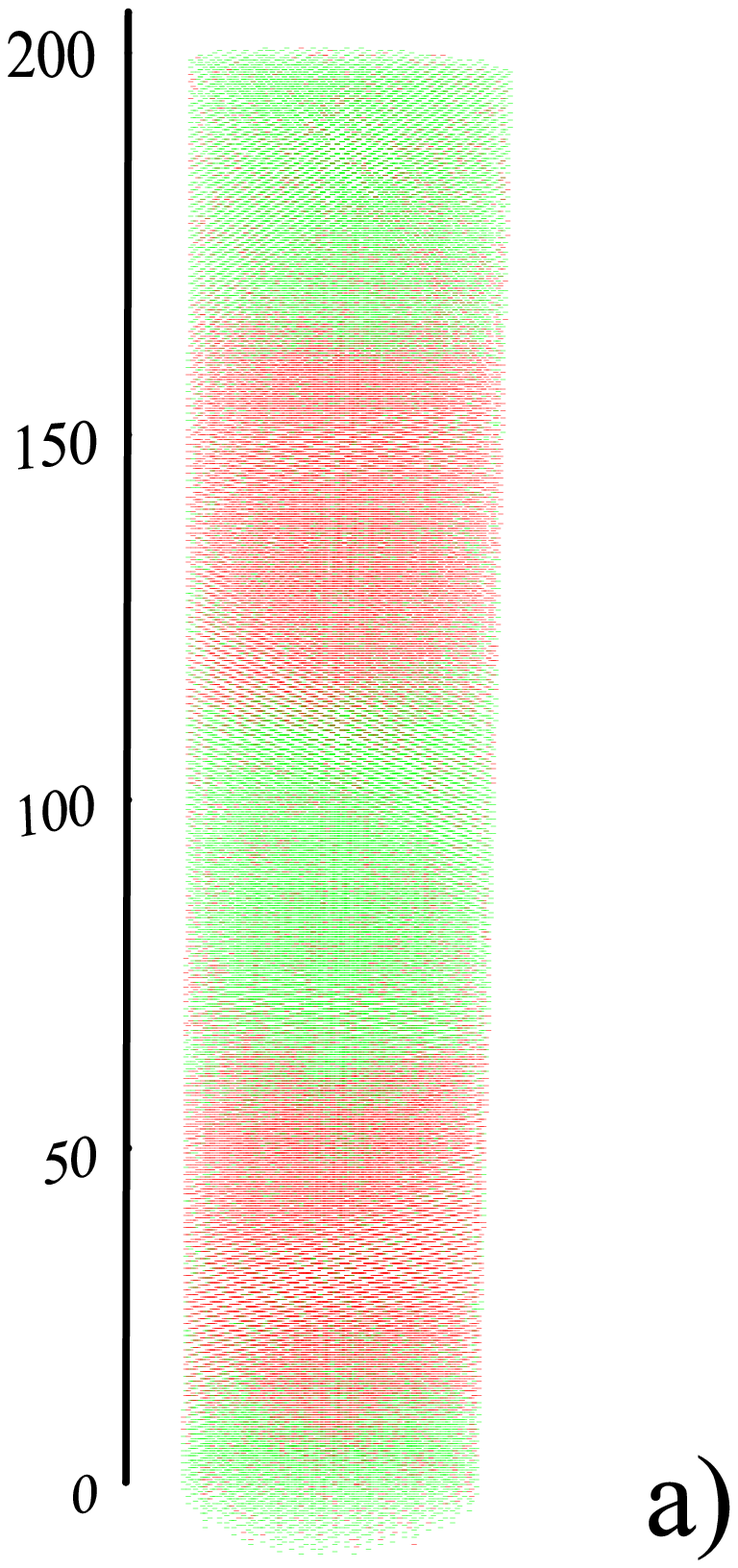}
\includegraphics[height=4.cm]{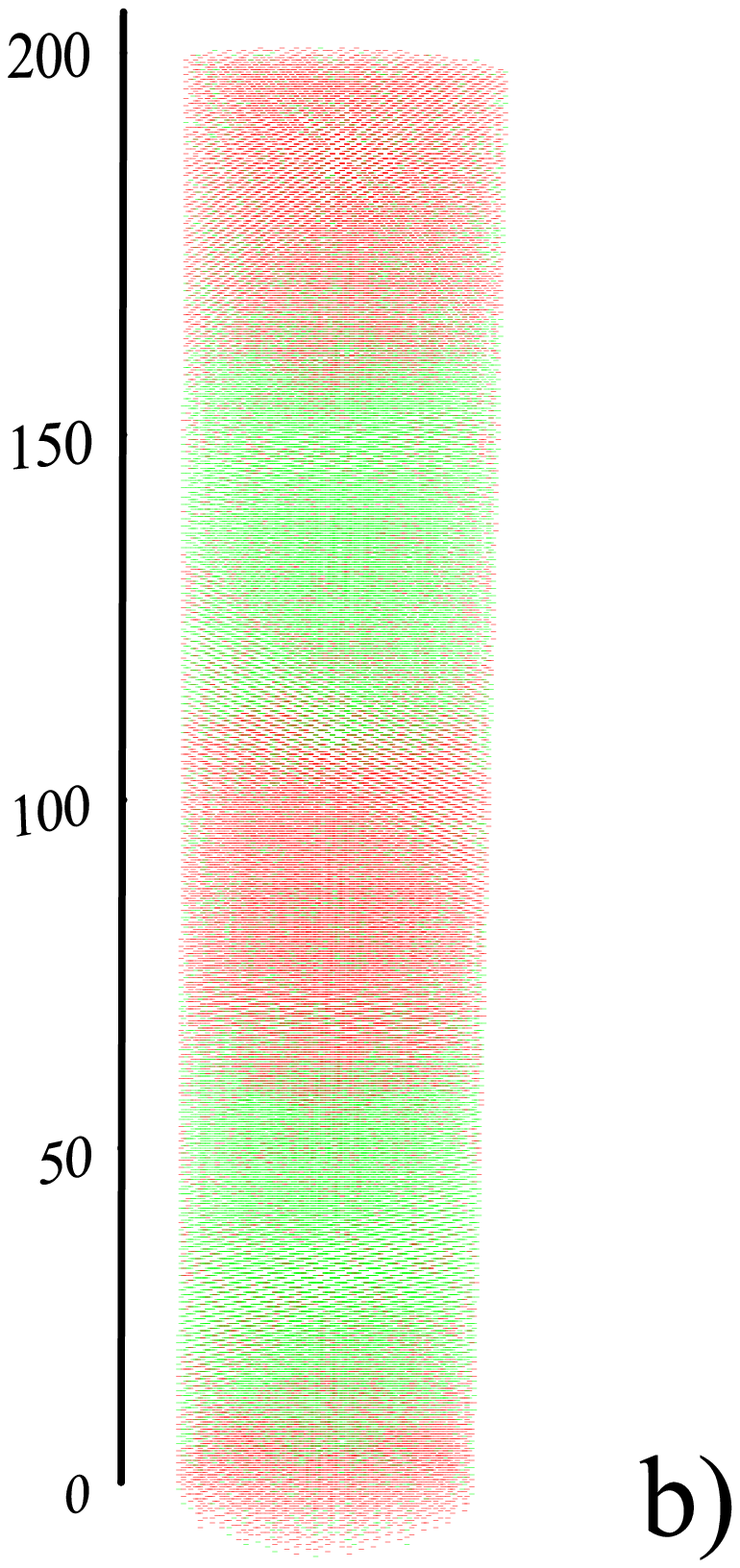}
\includegraphics[height=4.cm]{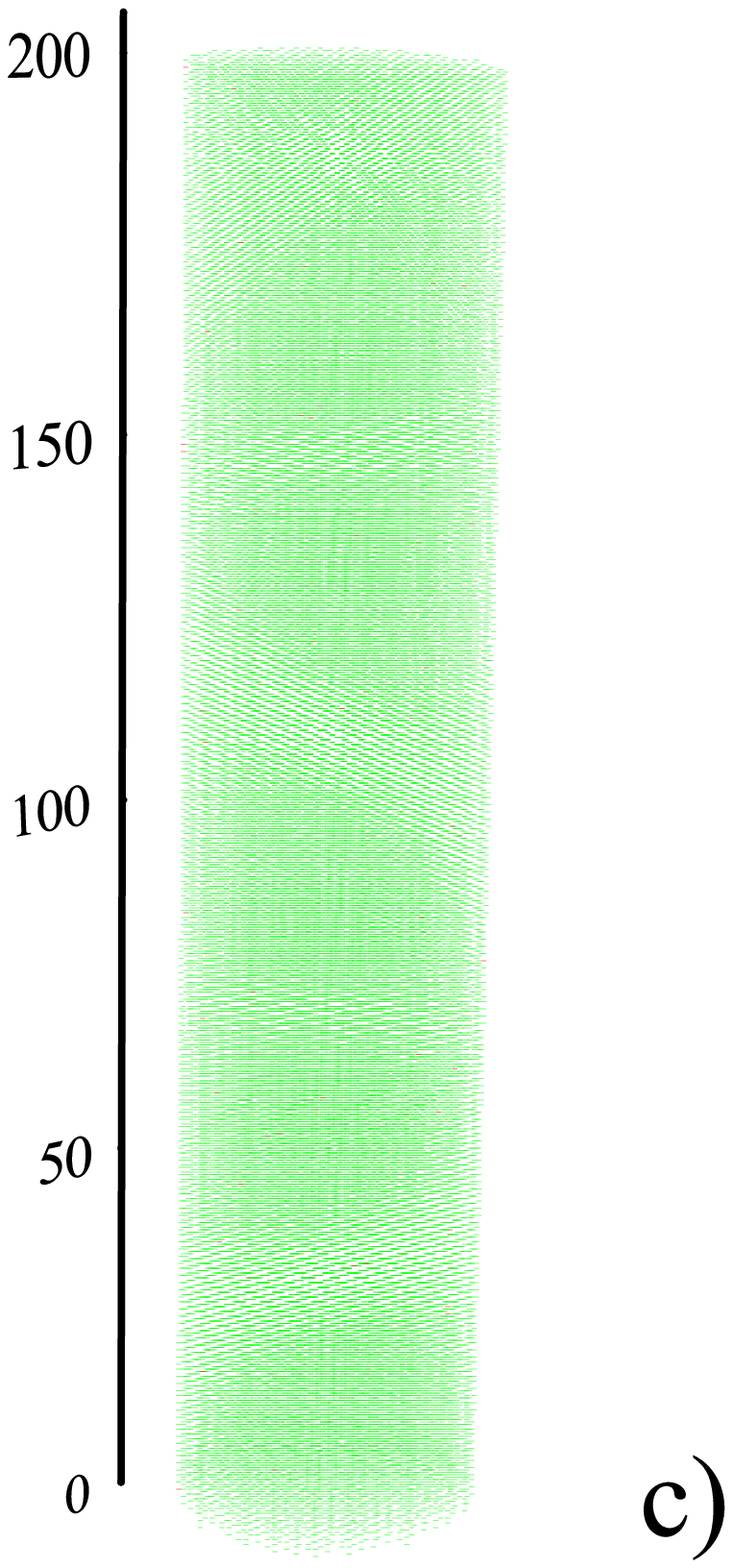}
\caption{(Color online) Coherent propagation of spin clusters of up (green dots) and down (red dots) spins,
swept by a propagating magnetic field wave. The spin snapshots are taken for two different values of the
Monte Carlo simulation time (a) $t=50100$ MCSS, (b) $t=50200$ MCSS for values of the reduced temperature $k_{B}T/J=4.0$,
amplitude $h_{0}/J=0.3$, wavelength $\lambda=100$ and period $\tau=100$ of the external
magnetic field. (c) corresponds to  spin snapshot at time $t=50100$ MCSS, for the same  system parameters
with (a) and (b), except from  considered value of  temperature  $k_{B}T/J=1.5$.}\label{Fig1}
\end{figure*}

The microscopic spin configurations of the present cylindrical nanowire system show that the system demonstrate
two different  dynamical  phases, which sensitively depend on the considered values of the
system parameters. Two alternate bands of spin  values $S=\pm1$  are found, and  they tend to
propagate along the long axis of the nanowire (namely along $z$ axis of nanowire),
in the high temperature regions. In Fig. \ref {Fig1}, we can see easily the coherent propagation of the
spin bands. These spin snapshots are taken for two different values of the MC  simulation time
(a) $t=50100$ MCSS, (b) $t=50200$ MCSS for values of the reduced temperature $k_{B}T/J=4.0$,
amplitude $h_{0}/J=0.3$ and  wavelength $\lambda=100$ of the external magnetic field.
The cylindrical nanowire  includes two full waves for the considered value
of $\lambda$.  Our MC simulation results also suggest
that a reduction in the value of temperature destroys the coherent propagating of spin bands of
the cylindrical nanowire system, leading to a spin-frozen or pinned phase \cite{Acharyya11}, as shown in
Fig. \ref{Fig1}(c) which is displayed for value of $k_{B}T/J=1.5$, with all other parameters of
propagating magnetic field remain the same. At the lower temperature regions, ferromagnetic spin-spin coupling
dominates against the  propagating magnetic field, hence, the most of the spins in the nanowire are frozen or
pinned to any one value of $S$. It is clear that these microscopic behaviors sensitively depend on
the studied system parameters.  We should note that these types of coherent propagations of spin
bands  and also spin-frozen phases have been recently observed in bulk materials under the
existence of  a propagating  magnetic field \cite{Acharyya11, Acharyya12, Acharyya13}.

\begin{figure*}[!h]
\center
\includegraphics[width=6.0cm]{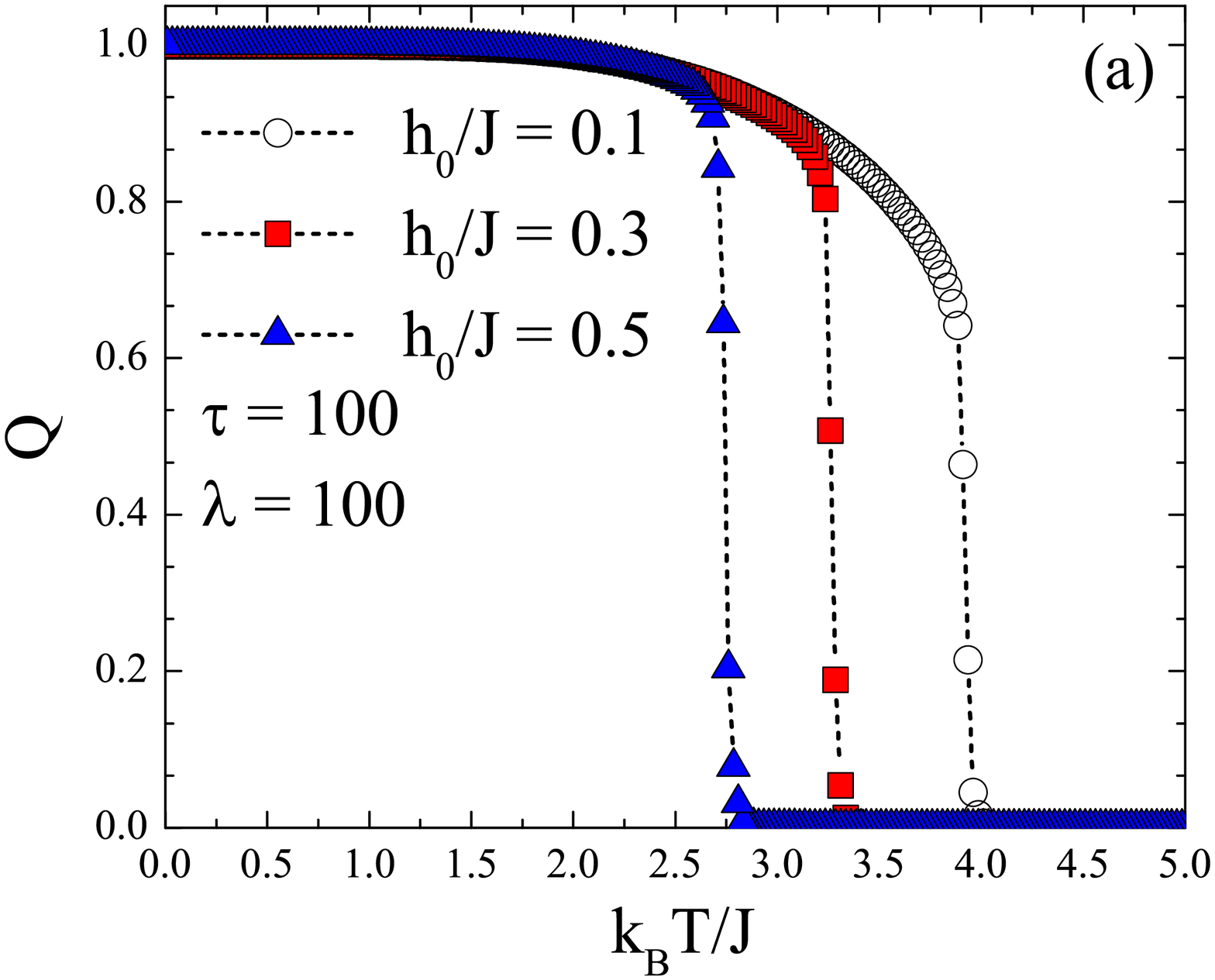}
\includegraphics[width=6.0cm]{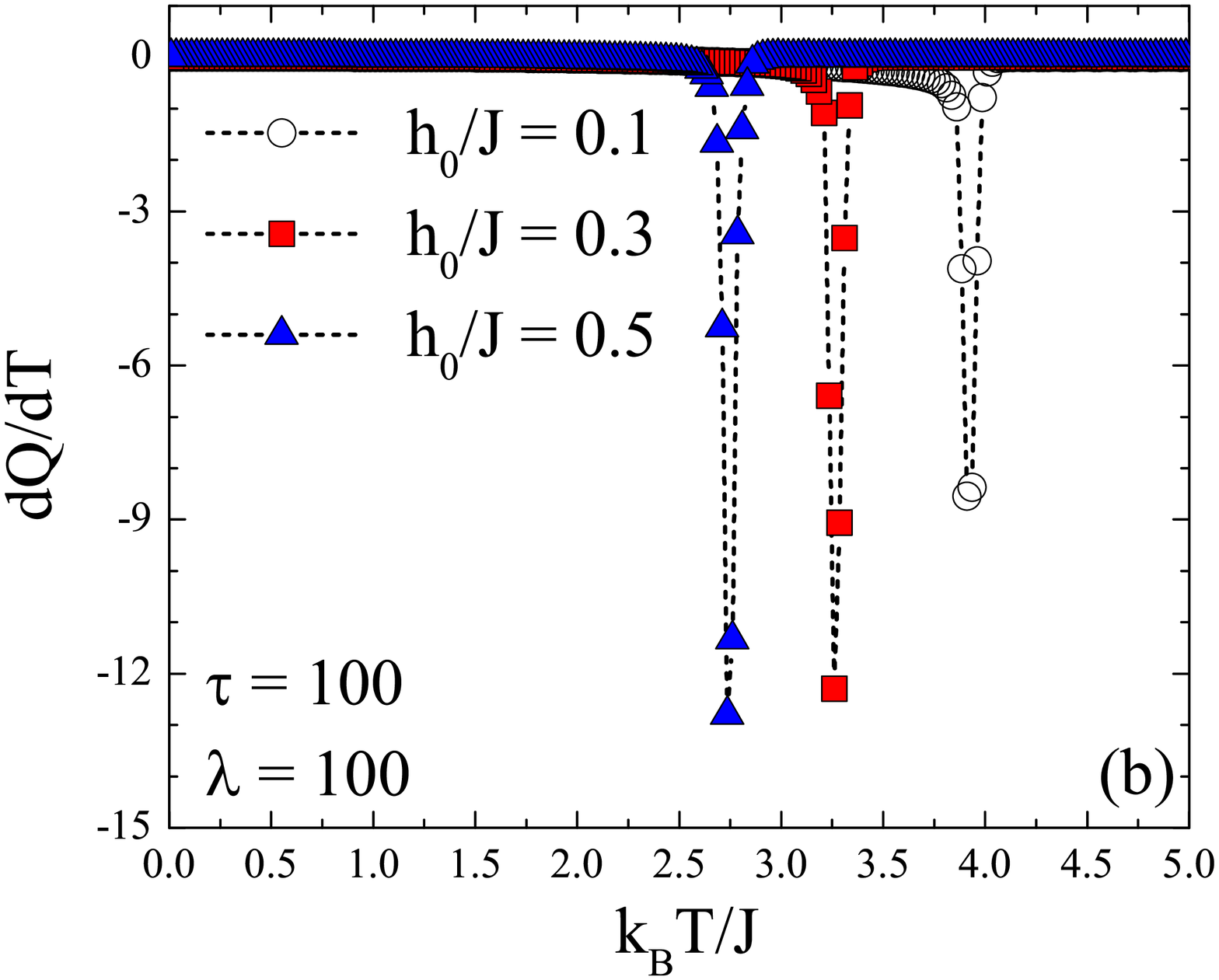}
\includegraphics[width=6.0cm]{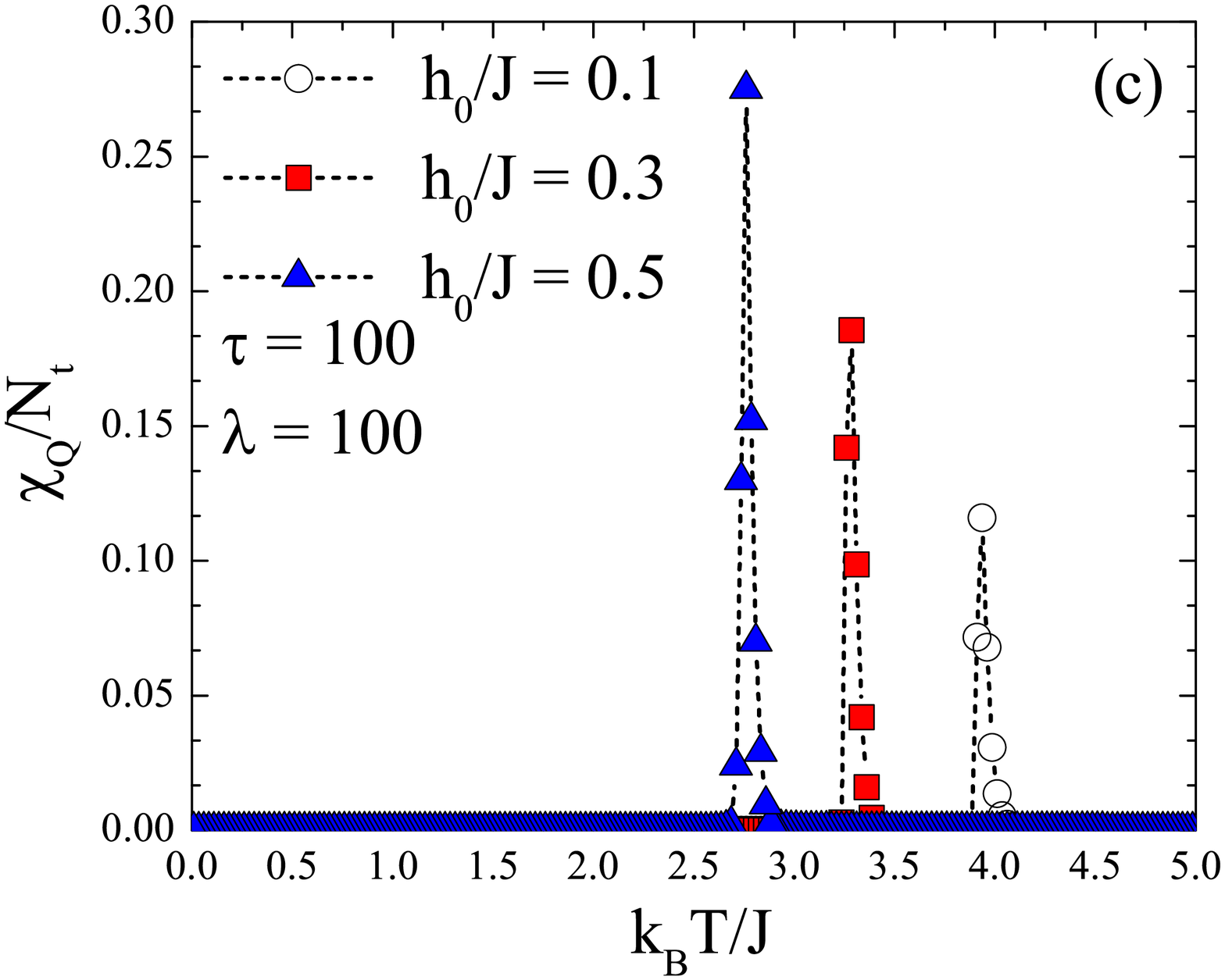}
\includegraphics[width=6.0cm]{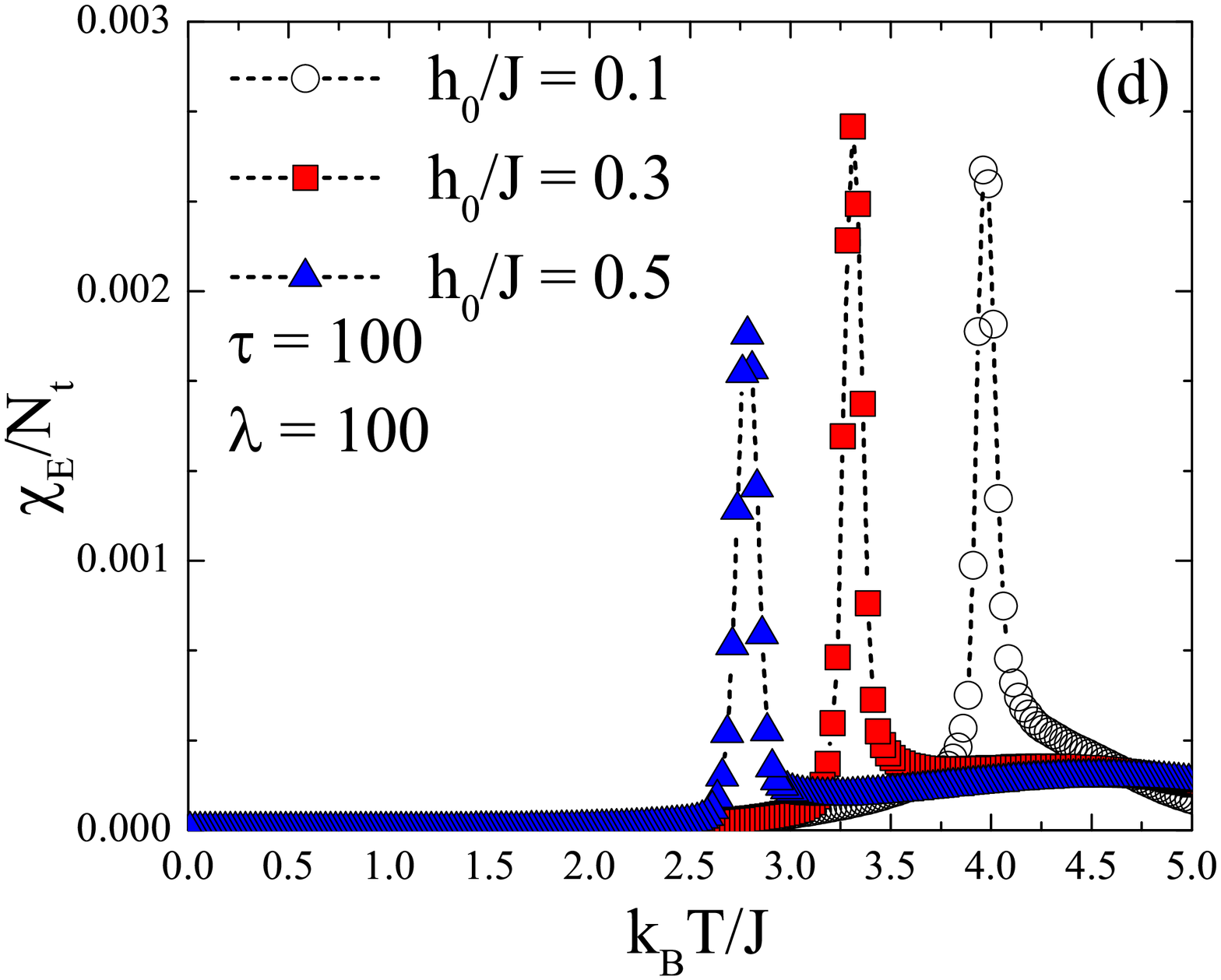}
\caption{(Color online) Thermal variations of the (a) dynamic order parameter $(Q)$, (b) $dQ/dT$, (c) variance
of $Q$ $(\displaystyle \chi_{Q}/N_{t})$ and (d) of $E$ $(\displaystyle \chi_{E}/N_{t})$ for different
values of reduced amplitude $h_{0}/J$ of the external field. The curves are depicted for values
of $\tau=100$ and $\lambda=100$.}\label{Fig2}
\end{figure*}

In Figs. \ref{Fig2}(a-d), we focus our attention on  the effects of the varying
field amplitudes  on the thermal  variations of the dynamic order
parameter $(Q)$, its derivative $dQ/dT$, variance  of $Q$ $(\chi_{Q})$ and of $E$ $(\chi_{E})$. The curves are depicted for
three values of the applied field amplitudes, i.e., $h_{0}/J=0.1, 0.3$ and $0.5$ with $\tau=100$
and $\lambda=100$. As shown in Fig. \ref{Fig2}(a), $Q$ gradually decreases starting
from its saturation value with increasing thermal energy, and it vanishes continuously at the critical temperature.
Dynamic phase transition point strongly depends on the component of the propagating magnetic field.
Our MC findings underline that an increment in value of the applied field  amplitude leads to a decrement in
the location of phase transition point. At the higher  temperature regions, $Q$ is zero and this
corresponds to a phase where the coherent propagation of spin bands exhibit.
In the lower temperature regions, the system exhibits dynamically ferromagnetic
phase, where $Q$ is nonzero. As discussed earlier, the microscopic spin snapshots show that the
system presents low temperature spin-frozen phase in these regions. Based
on these spin configurations of the cylindrical nanowire system,  it is possible to say that the
system undergoes a dynamical phase transition from a coherent spin  propagation phase  to a spin-frozen
phase, when the temperature is decreased starting  from a relatively higher value.
For a selected combination of the Hamiltonian parameters, the temperature dependencies
of $dQ/dT$ show a sharp dip while $\chi_{Q}$ and $\chi_{E}$ reveal a very sharp peak
indicating the existence of a  second order phase transition, as depicted
in  Figs. \ref{Fig2}(b-d). These peaks are found to shift to the lower
temperature values when the strength  of $h_{0}/J$ is increased.

\begin{figure*}[!h]
\center
\includegraphics[width=6.0cm]{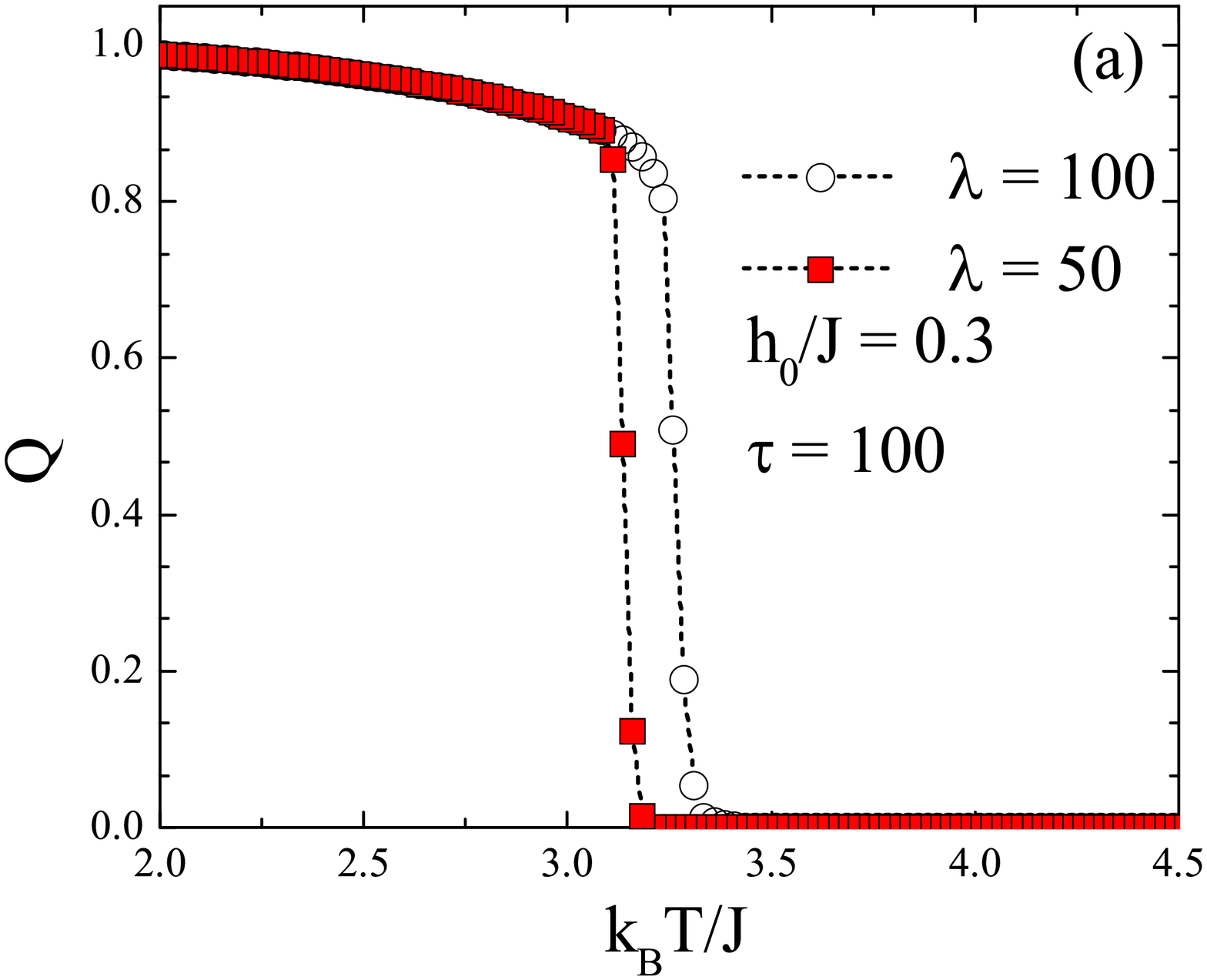}
\includegraphics[width=6.0cm]{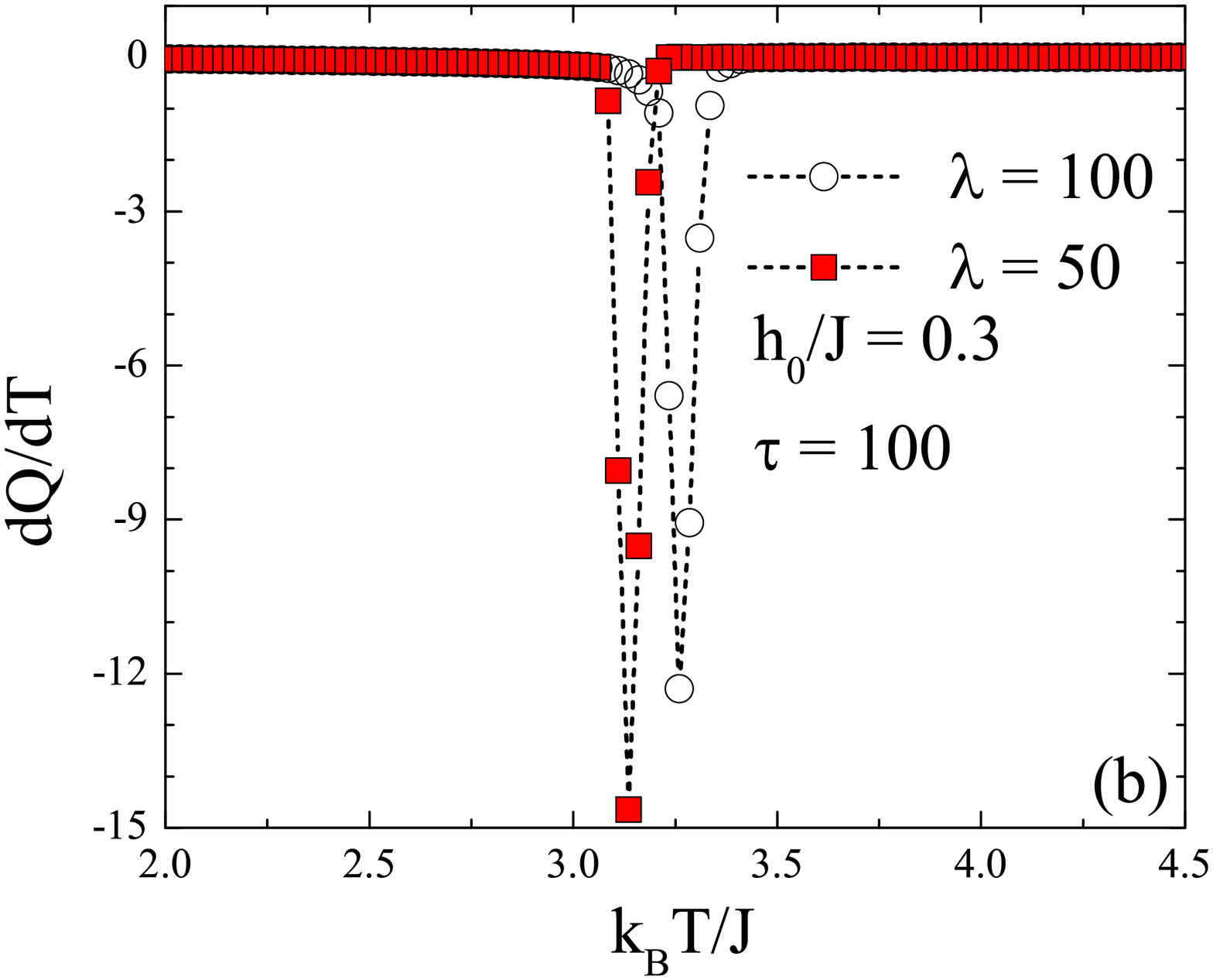}
\includegraphics[width=6.0cm]{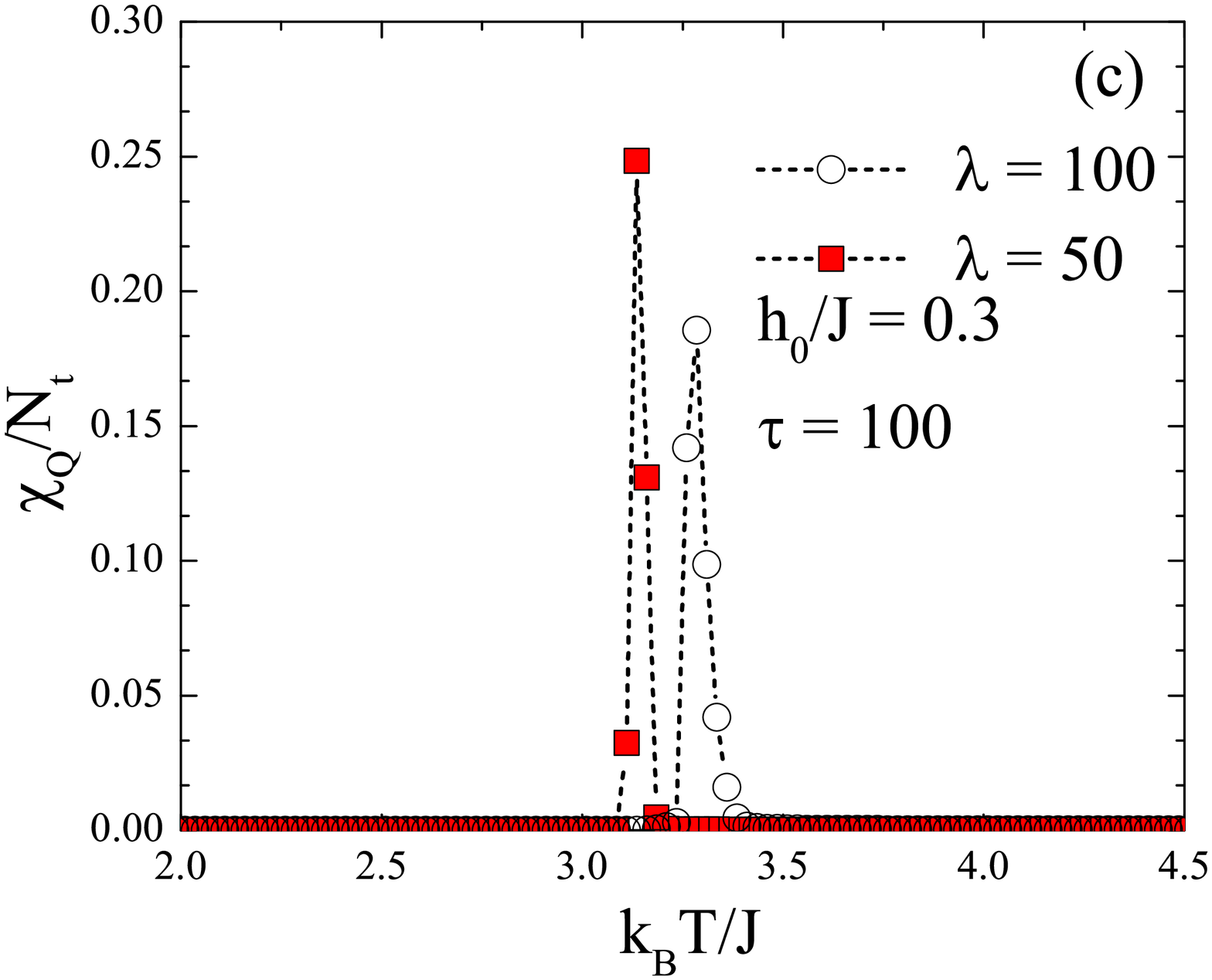}
\includegraphics[width=6.0cm]{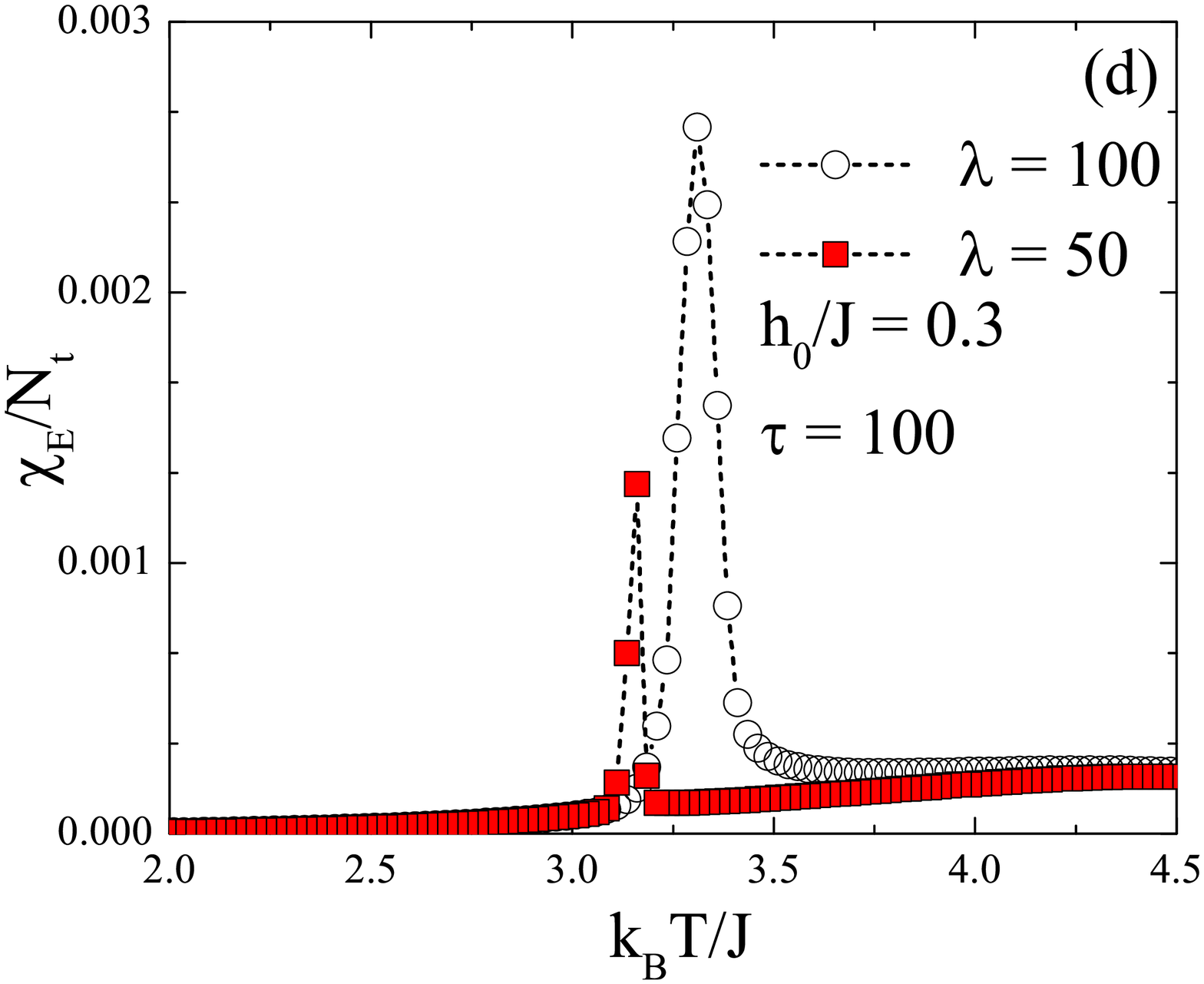}
\caption{(Color online) Temperature dependencies of the (a) $Q$, (b) $dQ/dT$,
(c) $\displaystyle \chi_{Q}/N_{t}$ and (d) $\displaystyle \chi_{E}/N_{t}$  at various
values of wavelength  $\lambda$ of  the external field. The curves are given for values of $\tau=100$
and $h_{0}/J=0.3$.}\label{Fig3}
\end{figure*}

In order to have a better understanding about the influences of the varying values of wavelength of
the propagating  field on  the system, we give the thermal variations of $Q$, $dQ/dT$, $\chi_{Q}$ and
also $\chi_{E}$ in Figs. \ref{Fig3}(a-d). The curves   are given for
two values of the wavelengths of the field, i.e., $\lambda=50$ and $100$ with $h_{0}/J=0.3$
and $\tau=100$. It is found  that dynamic phase transition
points are observed to depend on the selected value of $\lambda$ of the propagating
magnetic field wave. As seen in these figures, thermal variations of $Q$, $dQ/dT$, $\chi_{Q}$ and $\chi_{E}$
support that when the value of wavelength of the external field is increased, dynamically
ferromagnetic phase region gets wider. It may be noted here that  such types of observations
originating from the variation of the wavelength of the propagating magnetic field have been
found  in Ref. \cite{Acharyya12}, where kinetic Ising model on a $2D$ square lattice is exposed
to a propagating magnetic field.
\begin{figure*}[!h]
\center
\includegraphics[width=10cm]{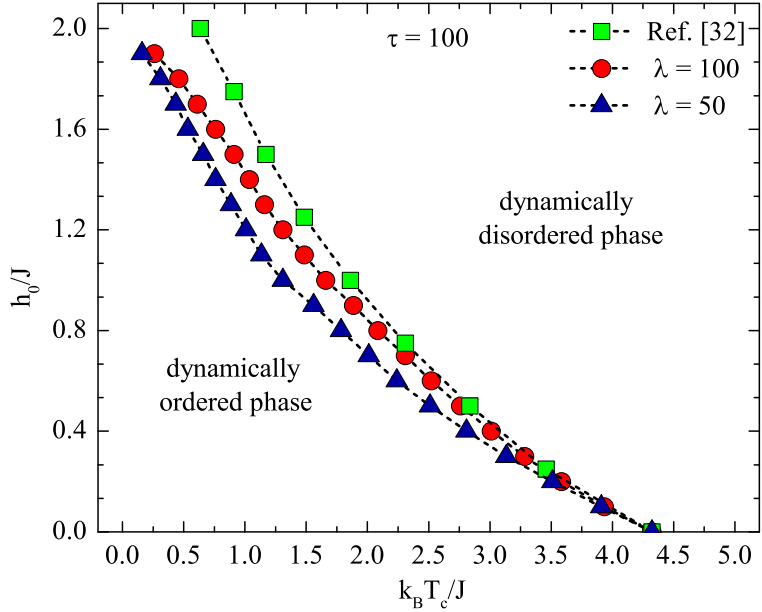}
\caption{(Color online) Dynamic phase diagram in the $\left(k_{B}T_{c}/J-h_{0}/J\right)$ plane
for the cylindrical nanowire system under the presence of a propagating magnetic
field. Different symbols correspond to the different values of applied field wavelengths: $\lambda=100$
(${\color{red}\bullet}$, red bullet) and  $\lambda=50$ (${\color{blue}\blacktriangle}$, blue triangle).
Here, ${\color{green}\blacksquare}$  (green square) symbols correspond to the phase transition points of the
spin-$1/2$ cylindrical nanowire system under a time dependent oscillating magnetic field,
which is uniform over space \cite{Yuksel2}, for the same system parameters used in this study.
Dynamic phase diagrams are given for value of $\tau=100.$}\label{Fig4}
\end{figure*}

In order to elucidate the influences of the applied field amplitude on the dynamic phase transition features
of the cylindrical nanowire system, we obtain the phase diagrams in $(k_{B}T_{c}/J-h_{0}/J)$ plane with two values of
the wavelength of the propagating magnetic field with $\tau=100$, in figure (\ref{Fig4}). We
note that dynamic phase transition points are deduced from the  peaks of the thermal  variations
of $dQ/dT$, $\chi_{Q}$ and $\chi_{E}$ curves. It is clear from the phase diagrams that when the
applied field amplitude is increased, the dynamic phase transition points are shifted to the
relatively lower temperature regions. The aforementioned behaviors seem to be independent of the applied
field wavelength $\lambda$. On the other hand, dynamic phase boundary line, which separates dynamically
ordered phases from disordered  phases, shrinks inward when the  strength of the  $\lambda$ of
the external field decreases. Recently, magnetic response of the spin-$1/2$ Ising cylindrical
nanowire system  to an oscillating magnetic field (uniform over space) has been
investigated by means of MC  simulation with Metropolis algorithm \cite{Yuksel2} for the same
system parameters with the present study. In order to make a comparison between propagating and oscillating magnetic fields,
the results of the reference \cite{Yuksel2} are added to the Fig. \ref{Fig4} (green squares).
It is obvious from the figure that there is no clear distinction between the magnetic field
sources for the small values of the applied field amplitudes, in the sense of
dynamic phase transition point. However, as the strength of $h_{0}/J$ is increased, dynamic
phase boundaries begin to seperate from each other. For considered values of the wavelengths of the
propagating magnetic field in this study, our MC simulation findings indicate that phase
transition points are always lower than  those of the critical points obtained in
reference \cite{Yuksel2}, especially at the higher applied field amplitude regions.

\section{Conclusions}\label{Conclusion}
To conclude, we study the nonequilibrium dynamics and phase transition features of the
ferromagnetic spin-$1/2$ cylindrical nanowire  under the existence of a  propagating magnetic
field. For this investigation, we use Monte Carlo simulation method  with single-site update
Metropolis algorithm. Based on the obtained results in this study,  it is possible to mention that
there are two types of dynamical states in the system,  depending on the considered system parameters.
The results can be summarized as follows:  In the higher temperature regions, two alternate
bands of spins are found, and they tend to coherently propagate along the long axis of the
nanowire. However, a decrement in  the value of temperature leads to the occurrence of
spin-frozen or spin pinned phase.  This phase generally occurs at the
relatively lower temperature regions, as in the case of the conventional bulk
systems under the influence of a propagating magnetic field \cite{Acharyya11, Acharyya12, Acharyya13}.

\noindent Additionally, we give the dynamic phase  diagram of the spin-$1/2$ cylindrical nanowire
in $(k_{B}T_{c}/J-h_{0}/J)$ plane  for two values of the wavelengths of the propagating magnetic
field.  Our MC simulation findings  clearly indicate that as the  strength of the  field amplitude is increased,
the phase transition points  tend to shift  to the  relatively lower temperature regions. In the
sense of wavelength of the field, it is found that dynamic  phase boundary line shrinks
inward in the related plane when  the value of  $\lambda$ of the external field decreases.
We also observed that the dynamic behavior  of the present system with a propagating magnetic
field exhibits quite different characteristics in comparison with the same system, but only in the
presence of a sinusoidally  oscillating magnetic field (uniform over space) \cite{Yuksel2}.

Very recently, it has been shown that frequency dispersion of the
nanocubic core/shell particle driven by an oscillating  field (uniform over space) can be
categorized into three groups, as in the case of the conventional bulk
systems \cite{Vatansever}. Keeping these in mind,
it would be interesting to investigate the influences of the propagating magnetic
field on the frequency dispersion of  dynamic loop area of the present system.
From the theoretical perspective, it could also be study the standing magnetic
field effects on the present system and (or) on the quenched disordered binary alloy
cylindrical nanowire \cite{ZVatansever}. We believe that such types of studies will
be beneficial to provide deeper understanding of physics underlying of nanoscale
materials driven by a time dependent magnetic field.

\section*{Acknowledgements}
The author is thankful to Muktish Acharyya from Presidency University for valuable discussions and suggestions.
The numerical calculations reported in this paper were
performed at T\"{U}B\.{I}TAK ULAKB\.{I}M (Turkish agency), High Performance and
Grid Computing Center (TRUBA	 Resources).

\end{document}